\documentclass[aps,twocolumn,showpacs,superscriptaddress]{revtex4-1}
\usepackage{amsmath,amssymb,amsfonts,graphics,graphicx,dcolumn,bm}

\newcommand{\cmp}
{\affiliation{Condensed Matter Physics Division, 
Saha Institute of Nuclear Physics, 1/AF Bidhannagar, Kolkata 700064, India.}}
\newcommand{\aalto}
{\affiliation{BECS, Aalto University School of Science, P.O.  Box 12200, FI-00076, Finland.}}
\newcommand{\isi}
{\affiliation{Economic Research Unit, Indian Statistical Institute, 203 B. T. Road, Kolkata 700108, India.}}
\newcommand{\bu}
{\affiliation{Department of Economics, Boston University, 270 Bay State Road, Boston, MA-02134, USA.}}

\begin{document}

\title
{Zipf's law in city size from a resource utilization model}

\author{Asim Ghosh}
\email[Email: ]{asim.ghosh@saha.ac.in}
\cmp
\author{Arnab Chatterjee}%
\email[Email: ]{arnabchat@gmail.com}
\aalto \cmp
\author{Anindya S. Chakrabarti}%
\email[Email: ]{anindya@bu.edu}
\bu
\author{Bikas K Chakrabarti}%
\email[Email: ]{bikask.chakrabarti@saha.ac.in}
\cmp \isi

\begin{abstract}
We study a resource utilization scenario characterized by intrinsic fitness.
To describe the  growth and organization of different cities, we consider a model for resource utilization 
where many restaurants compete, as in a game,
to attract customers
using an iterative learning process.
Results for the case of restaurants with uniform fitness are reported.
When fitness is uniformly distributed, it gives rise to a Zipf law
for the number of customers.
We perform an exact calculation for the utilization fraction for 
the case when choices are made independent of fitness.
A variant of the model is also introduced where the fitness can be treated 
as an ability to stay in the business. When a restaurant loses customers, its
fitness is replaced by a random fitness. 
 The steady state fitness distribution is characterized by a power law,
 while the distribution of the  number of customers 
 still follows the Zipf law, implying the robustness of the model.
Our model serves as a paradigm for the emergence of Zipf law in city size distribution.
\end{abstract}
\maketitle

\section{Introduction}
The complexity of interactions in human societies have 
produced various emergent phenomena~\cite{Castellano:2009,Sen:2013}, 
often characterized by broad distributions of different quantities. 
One of the interesting consequences of economic growth is urban agglomeration.
A striking example of agglomeration
is expressed as a broad distribution 
for urban entities -- city sizes, given by their population, and 
first reported by Auerbach~\cite{auerbach1913}.
Known to be the Zipf law~\cite{Zipf1949},
city sizes follow
a simple distribution law: the rank $k$ of a city
with population $s$ goes as $s_k \sim 1/k^{\gamma}$ with the Zipf exponent $\gamma \approx 1$
holding true for most societies and across time.
However, variations to this structure have also been observed for 
countries like China or the former USSR countries~\cite{Gangopadhyay2009,Benguigui2007}. 
The probability density of city sizes follow from above,
again a power law: $P(s) \sim s^{-\nu}$ ($\nu > 0$).
The exponents of the Zipf plot $\gamma$ and that corresponding to the 
probability density $\nu$ are related as
$\nu=1+\frac{1}{\gamma}$~\cite{Newman2005}.

Several studies attempted to derive the Zipf's law theoretically for city-size distributions,
specifically for the case $\gamma=1$.
Gabaix~\cite{Gabaix1999} argued that if cities grow randomly at the same expected growth rate
and the same variance, the limiting distribution will converge to Zipf's law.
In a similar approach, resulting in 
diffusion and multiplicative processes, produced intermittent spatiotemporal structures~\cite{zanette1997role}.
Another study used shocks as a result of migration~\cite{Marsili1998}.
In Ref~\cite{Gabaix1999} however, differential population growth resulted from migration.
Some simple economics arguments showed that the expected urban
growth rates were identical across city sizes and variations were random normal deviates, 
and the Zipf law with exponent unity follows naturally.

Zipf law has also been observed for 
firm sizes~\cite{Axtell2001}, 
income distribution of companies~\cite{Okuyama1999},
firm bankruptcy~\cite{Fujiwara2004}, etc.

Cities are characterized by their economic output, wealth, 
employment, wages, housing condition, crime, transport 
and various other amenities~\cite{Bettencourt2007},
and can also be quantitatively evaluated and ranked using various indices
(e.g., Global City index~\cite{cityindex}).
Historically, cities have seen birth, growth, competition, migration, decline
and death, but over time, the ranking of cities
according to size is claimed to be following a Zipf law irrespective of 
time~\cite{batty2006rank}.
While people choose to live in cities deciding on different factors,
and compete to make use of the resources provided by the cities, the 
migration of population across cities to adjust for resources~\cite{beaudry2014spatial}
also plays an important role in the city growth/decay dynamics.

One of the toy models to study resource utilization~\cite{Chakraborti2013review} 
is the Kolkata Paise Restaurant (KPR)~\cite{Chakrabarti2009,Ghosh2010}
problem, which is similar to various adaptive games (see \cite{Challet2004}). In the simplest version, 
$N$  agents (customers) simultaneously choose between equal number  $R$ ($=N$) of restaurants,
each of which serve only one meal every evening (generalization to any other number is trivial). 
Thus, showing up in a restaurant with more people 
means less chance of getting food.
The \textit{utilization} is measured by the fraction
of agents $f$ getting food or equivalently, by measuring its complimentary quantity: 
the fraction of meals wasted ($1-f$), since some restaurants do not get any customer at all.
A fully random occupancy rule provides a benchmark of $f= 1-1/e \approx 0.63$, while
a crowd-avoiding algorithm~\cite{Ghosh2010} improves the utilization to around $0.8$.
It was also seen that varying the ratio of the number of agents to the number of restaurants $(N/R)$
below unity, one can find a phase transition between a `active phase' characterized
by a finite fraction $\rho_a$ of restaurants with more than one agent, and an `absorbed phase',
where  $\rho_a$ vanishes~\cite{Ghosh2012}. The same crowd avoiding strategy was adapted
in a version of the Minority Game~\cite{Challet2004} which provided the extra information
about the crowd and one could achieve a very small time of convergence to the steady state, 
in fact $O( \log \log N)$~\cite{Dhar2011}. Another modification to this problem~\cite{Biswas2012}
showed a phase transition depending on the amount of information that is shared.
The main idea for the above studies was to find simple algorithms that lead to a state of 
maximum utilization in a very short time scale, using iterative learning. 

But in reality, resources are never well utilized, and in fact, socio-economic
inequalities
are manifested in different forms, among which the inequalities in
income and wealth~\cite{Yakovenko2009,Chakrabarti2013book} are the most prominent
and quite well studied.
While empirical data gave us an idea of the form of the distribution 
of income and wealth, various modeling efforts have supplemented them
to understand why such inequalities appear.
One of the successful modeling attempts used the kinetic theory of gases~\cite{Dragulescu2000},
where gas molecules colliding and exchanging energy was mapped into agents
coming together to exchange their wealth, obeying certain rules~\cite{Chatterjee2007}.
Using savings as a parameter one can model the entire range of the 
income/ wealth distribution. 
In the models, a pair of agents agree to trade and each save a fraction $\lambda$ of their 
instantaneous money/wealth and performs a random exchange of the rest at each trading step.
The distribution of wealth $P(m)$ in the steady state matches well with the characteristic empirical data.
When the saving fraction $\lambda$ is fixed, i.e., for homogeneous agents (CC model hereafter)~\cite{Chakraborti2000}, 
$P(m)$ resemble Gamma distributions~\cite{Patriarca2004}.
When $\lambda$ is distributed random uniformly in $[0,1)$ and quenched, (CCM model hereafter),  i.e., for heterogeneous agents,
one gets a Pareto law for the probability density $P(m) \sim m^{-\nu}$ with exponent $\nu=2$~\cite{Chatterjee2004,Chatterjee2007}.
This model uses preferential attachment~\cite{barabasi1999emergence} with socio-economic ingredients.

In this paper, we connect the setting of the KPR problem with kinetic 
exchange models of wealth distribution.  Customers migrate across restaurants depending
on their satisfaction, where the saving fraction of agents in the
kinetic exchange models of wealth distributions correspond to the 
fitness of the restaurants.
This  serves as a model for city growth and organization, where the cities correspond to restaurants
and the city population to the customers, who choose to stay or migrate according to the fitness
of the cities.

In Sec.~\ref{sec:2}, we define our model in which each restaurant has an inherent \textit{fitness}
which keeps agents from going away to other places. In Sec.~\ref{sec:3}, we perform calculations for size distributions
as well as utilization
fraction for cases with uniform and distributed fitness parameter. In a modified version of the model
with fitness, the results are shown to be robust.
In Sec.~\ref{sec:4}, we discuss some empirical evidences. We conclude with
summary and discussions in Sec.~\ref{sec:5}.

\section{Model}
\label{sec:2}
In the usual KPR framework of $N$ agents and $R$ restaurants, we take here in the following 
$R=N$ for the sake of simplicity.
We assume that each restaurant $i$ has a characteristic \textit{fitness} $p_i$
drawn from a distribution $\Pi(p)$. The entire dynamics of the agents is defined by $p$.
The concept of time is similar in the case of cities in the sense that people make choices
at a certain time scale.
Agents visiting a restaurant $i$ on a particular evening $t$ return on the next evening $t+1$ with 
probability $p_i$, or otherwise go to any other randomly chosen restaurant.
We consider the dynamics of the agents to be simultaneous.

In terms of cities,  we can restate the model as follows: every city has some \textit{fitness}  
and initially people are randomly distributed among the cities. At any point of time, some people 
will be satisfied in a city and others will not be satisfied by its services. 
According to our model, the unsatisfied  people will shift randomly to any other cities. 
The same dynamics happens for other cities too. Therefore at every time step (which can of the order of days or months) 
cities may lose some people and may also gain some people.  
We consider different types of \textit{fitness} distribution and observe the population distribution for the cities.

The \textit{fitness} parameter above is a proxy for a generic city index~\cite{cityindex}, 
which can be any intrinsic property such as the measure of wealth, economic power, competitiveness, resources, 
infrastructure etc. or a combination of many of these.
It is important to note at this point that we are using the restaurant model (KPR) paradigm
to model the distribution of sizes of urban agglomerations (cities), where migration 
between cities is modeled by the movement of agents across restaurants.

In order to measure utilization, we further assume that the restaurants prepare as many meals on a particular evening
as there were customers on the previous evening. 
Thus the restaurants learn to minimize their wastage.
The wastage $1-f$ given by the unused meals, and the utilization fraction $f$ can thus be computed.
Note that the utilization fraction $f$ here is different from that used earlier in Refs.~\cite{Chakrabarti2009,Ghosh2010,Ghosh2012,Biswas2012}
in the sense that restaurants here `learn' also to adjust the size of their services according to their past experience.

\section{Results}
\label{sec:3}
\subsection{Distribution of sizes}
\label{subsec:size}
\begin{figure}[t]
 \includegraphics[width=8.7cm]{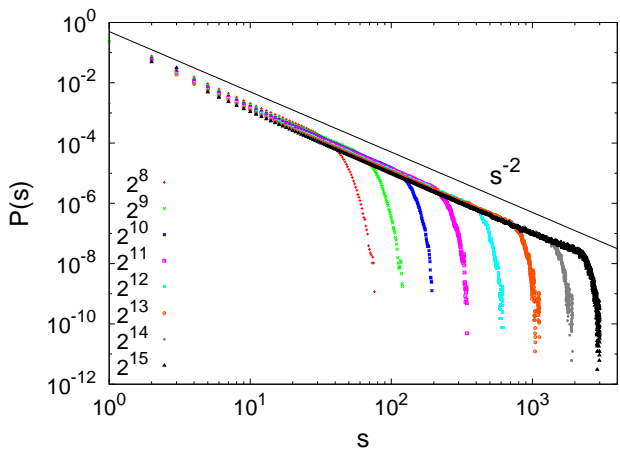}
 \caption{(Color
  online) 
 The probability density $P(s)$ for fraction of restaurants with $s$ agents.
 The data is shown for different system sizes $N=2^8, 2^9, 2^{10}, 2^{11}, 2^{12}, 2^{13}, 2^{14}, 2^{15}$.
 The power law exponent is compared with $s^{-2}$.
 }
 \label{fig:Ps}
\end{figure}

Let us consider the case when $p_i$ is uniformly distributed in $[0,1)$,
i.e, $\Pi(p)=1$. In practice, we use a natural cutoff for $p$
as $1-1/N$. The probability density of the number of agents $s$
at a particular restaurant $P(s)$ has a broad distribution, and in fact
a power law for most of its range, but has a prominent exponential cutoff:
\begin{equation}
 P(s) \sim s^{-\nu} \exp(-s/S),
\end{equation}
where $S$ is a constant which determines the scale of the cutoff.
The exponential cutoff is an artifact of the upper cutoff in $\Pi(p)$.
The power law exponent is $\nu=2.00(1)$ as measured directly from the fit 
of the numerical simulation data (Fig.~\ref{fig:Ps}).

Let $a_i(t)$ denote the number of customers on the evening $t$ in the restaurant $i$ 
characterized by fitness $p_i$ in the steady state. So, $\sum_i a_i(t) =N$.
Let $n^{\prime}$ denote the average number of agents on any evening who are choosing
restaurants randomly. Then, for a restaurant $i$, $a_i(t) p_i$ agents are returning
to restaurant $i$ on the next evening, and an additional $n^{\prime}/N$ agents on the average 
additionally come to that restaurant. This gives
\begin{equation}
 \overline{a_i(t+1)} = \overline{a_i(t)} p_i + n^{\prime}/N,
\end{equation}
where $\overline{a_i}$ would now denote the average quantity.
In the steady state, we have $\overline{a_i(t+1)}= \overline{a_i(t)} = \overline{a_i}$ and hence 
\begin{equation}
 \overline{a_i} (1-p_i) = \frac{n^{\prime}}{N} \label{eq:const}
\end{equation}
giving
\begin{equation}
 \overline{a_i} = \frac{n^{\prime}}{N} \frac{1}{1-p_i}.
 \label{eq:const1}
\end{equation}
These calculations hold for large $p_i$ (close to $1$) which give large values of
$a_i$ close to  $\overline{a_i}$.
Thus, for all restaurants, 
\begin{eqnarray}
 \sum_i \overline{a_i} &=& N = \frac{n^{\prime}}{N} \sum_i \frac{1}{1-p_i} \nonumber \\
 \Rightarrow n^{\prime} &=& \frac{N^2}{\sum_i \frac{1}{1-p_i}}.
\end{eqnarray}
Now, let us consider a case of $\Pi(p)=1$, where $p_i = 1-i/N$ for $i=1,2,\ldots,N$. Thus,
\begin{equation}
 n^{\prime} = \frac{N}{\sum_i \frac{1}{i}} \approx \frac{N}{\ln (N+1)}
 \label{eq:log}
\end{equation}
for large $N$. We numerically computed $P(s)$ for this particular case
and the computed value of the cutoff in $P(s)$ which comes from the largest value
of $p_i$ which is $p_1=1-1/N$, and it agrees nicely with our estimated Eq.~\ref{eq:log}.

\begin{figure}[t]
 \includegraphics[width=8.7cm]{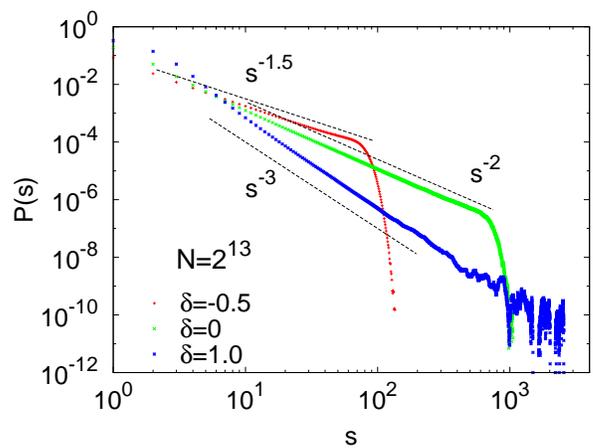}
 \caption{(Color
  online) 
 The probability density $P(s)$ for fraction of restaurants with $s$ agents,
 for different distributions $\Pi(p) = (1+\delta)(1-p)^{\delta}$, with $\delta=-0.5,0,1.0$.
 The power law exponents agree with $\nu=2+\delta$.
 The data are shown for $N=2^{13}$.
 }
 \label{fig:delta}
\end{figure}

%
Following Ref.~\cite{Mohanty2006}, one can derive the form of the size distribution $P(s)$ easily.
Since, R.H.S. of Eq.~(\ref{eq:const}) is a constant ($=C$, say), $dp = da/a^2 = ds/s^2$, since
$a_i$ being the number of agents in restaurant $i$ denotes nothing but the size $s$.
An agent with a particular fitness $p$ ends up in a restaurant of
characteristic size $s$ given by Eq.~(\ref{eq:const}), so that one can relate
$\Pi(p) dp = P(s) ds$. Thus,
\begin{eqnarray}
 P(s) &=& \Pi(p) \frac{dp}{ds} = \frac{\Pi \left(1 - \frac{C}{s} \right)}{s^2}.
 \label{eq:derive}
\end{eqnarray}
Thus, for an uniform distribution $\Pi(p)=1$, $P(s) \sim s^{-2}$ for large $s$.
It also follows that for $\Pi(p) = (1+\delta)(1-p)^{\delta}$, one should get
\begin{equation}
 P(s) \sim s^{-(2+\delta)}, \; \; \textrm{with} \; -1 < \delta < \infty.
 \label{eq:2+d}
\end{equation}
Thus $\nu$ does not depend on any feature of $\Pi(p)$ except on the nature of this function near $p=1$, i.e., the value of $\delta$,
giving $\nu = 2+\delta$.

\begin{figure*}[t]
 \includegraphics[width=8.7cm]{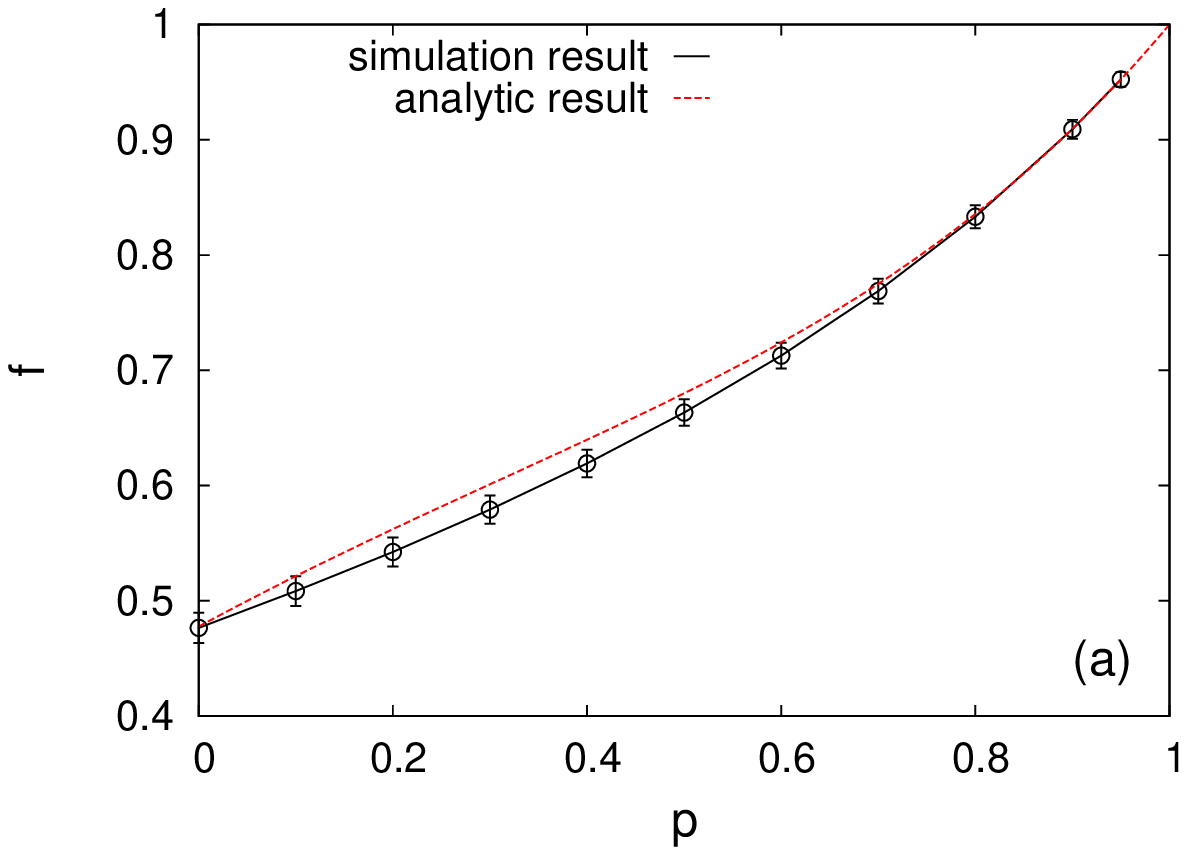}
 \includegraphics[width=8.7cm]{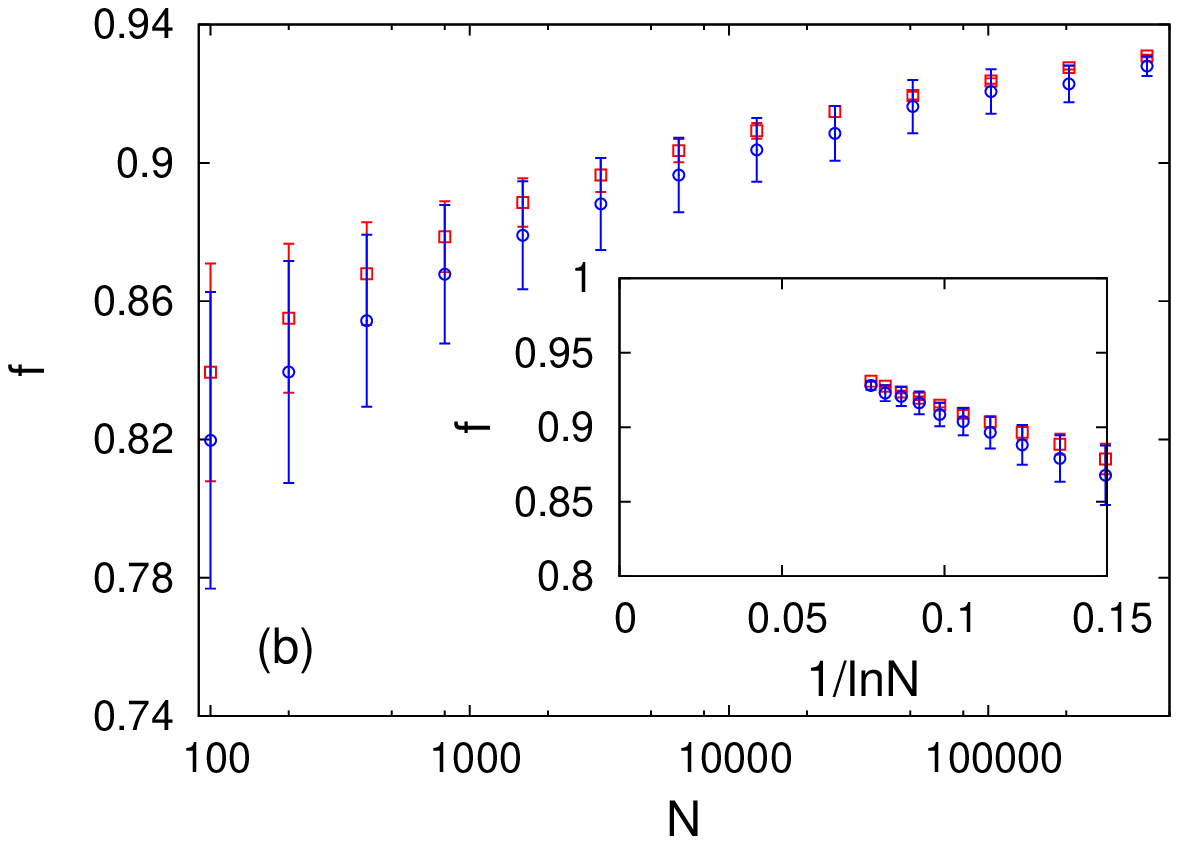}
 \caption{(Color
  online) Utilization fraction $f$ in the model (a)  with  fixed fitness $p$.
 The data is shown for $N=1024$. The dotted line gives the analytical estimate.
 (b) $f$  for the case of uniformly distributed $p$ i.e., for $\Pi(p) =1$: 
 $p_i =1-i/N$ (red squares) and uniformly random (blue circles) for various system sizes $N$. 
 The inset shows the variation of $f$ with $1/\ln N$. Error bars are also shown.
 }
 \label{fig:util}
\end{figure*}
Eq.~\ref{eq:2+d} can also be derived in an alternative way.
The fraction of redistributed people $n^\prime/N$  
choose $N$ restaurants randomly, and thus its distribution
is Poissonian of some parameter $c$. 
The stationary distribution of $a_i$ is also Poissonian
of parameter $c/(1-p_i)$.
The average distribution over $p_i$ can hence be computed exactly as 
\begin{eqnarray}
P(a)&=&\frac{\delta+1}{a!}\int_0^{1-\epsilon}dp(1-p)^{\delta}\left(\frac{c}{1-p}\right)^a \exp \left(-\frac{c}{(1-p)} \right)\nonumber \\
    &=& \frac{(\delta+1)c^{\delta+1}}{a!} \int_c^{c/\epsilon} u^{a-2-\delta} \exp(-u)du,
    \label{eq:pa}
\end{eqnarray}
where $\epsilon=1/N^{\frac{1}{\delta+1}}$ is a cutoff, in particular necessary for $\delta=0$, and
\begin{equation}
c = 
\left\{ \begin{array}{lll}
\frac{1}{(\delta+1)} \int_0^{1-\epsilon}dp(1-p)^{\delta-1} &= \frac{\delta}{(\delta+1)}, & \textrm{if} \ \delta>0,\\
\frac{1}{\log(1/\epsilon)} &=\frac{1}{\log(N)}, & \textrm{if} \ \delta=0.
\end{array}\right.
\end{equation}
If the bounds of the above integral (Eq.~\ref{eq:pa}) can be put
respectively to $0$ and $\infty$, for an intermediate range of $a$ one gets
\begin{eqnarray}
P(a) & \sim& \frac{(a-2-\delta)!}{a!} \nonumber \\
     &\sim& a^{-(2+\delta)}, \; \; \textrm{when} \; a \gg 2+\delta,
\end{eqnarray}
which leads to the result below Eq.~\ref{eq:2+d}, while Eq.~\ref{eq:2+d} is only valid
for large $a$.

In Fig.~\ref{fig:delta} we compare the numerical simulation results 
for $\Pi(p) = (1+\delta)(1-p)^{\delta}$ and indeed find the agreement $\nu=2+\delta$.

At this point, it is worthwhile to mention the case when restaurants have the same
fitness, $p_i = p$ $\forall i$. The $p=0$ case is trivial and same as
our random benchmark. The size distribution $P(s)$ is Poissonian: $P(s)=\frac{1}{s!}\exp(-1)$. 
For  $0 < p < 1$, $P(s)$ does not show any difference, except in the largest values of $s$. 
Trivially, the $p=1$ case has no dynamics.
This is strikingly different from the CC model~\cite{Chakraborti2000}, where
the wealth distribution $P(m)$ resembles Gamma distributions~\cite{Patriarca2004}, with the maxima
for $\lambda=0$ at $m=0$ monotonically going to $m=1$ for $\lambda \to 1$, $m$ being
calculated in units of average money per agent.
However, in the limit of $g=N/R \gg 1$ (continuum limit), 
the above models will reproduce results of CC and CCM.

\subsection{Utilization}
\label{subsec:util}
We further assume that the restaurants prepare as many meals on a particular evening
as there were customers on the previous evening.
We define utilization fraction $f$ as the average fraction of agents getting food.
Thus, formally,
\begin{equation}
 f = \left \langle \overline{ \frac{1}{N} \sum_i \min [a_i(t),a_i(t+1)]} \right \rangle,
 \label{eq:util}
\end{equation}
where the bar means time average in the steady state and $\langle \ldots \rangle$ means ensemble average.
Thus, Eq.~\ref{eq:util} computed in the steady state will give the steady state
value of utilization $f$.

Let us consider the case when the agents choose restaurants randomly.
The utilization fraction $f$ is about $0.476(5)$ as computed from
numerical simulations. We can provide an analytical argument for this.

The probability of finding a restaurant with exactly $m$ agents is given by
\begin{equation}
 \pi(m) = \frac{1}{m!} \exp(-1).
\end{equation}
In the steady state, $\pi(m)$ fraction of restaurants each provide $m$ meals.
Then the fraction of agents not getting food can be calculated exactly, and is given by
\begin{eqnarray}
 1-f = \pi(0) &+& \pi(1) [1.\pi(2)+ 2.\pi(3) + \ldots] \nonumber \\
     &+&        \pi(2) [1.\pi(3)+ 2.\pi(4) + \ldots]  \nonumber \\
     &+&        \pi(3) [1.\pi(4)+ 2.\pi(5) + \ldots] + \ldots \\
    = \pi(0) &+& \sum_{r=1}^{\infty} \sum_{r^{\prime}=r+1}^{\infty} \pi(r) \pi(r^{\prime}) (r^{\prime} - r).
\label{eq:series}
\end{eqnarray}
Eq.~(\ref{eq:series}) can be computed to any degree of accuracy.
The series for its first four terms, i.e., keeping upto $r=3$, gives
$1-f = \frac{1}{e} + \frac{1}{e^2} + \frac{1}{2e^2} (3-e) + \frac{1}{6e^2} \left(\frac{11}{2} -2e \right)  \approx 0.523$.
Thus, $f \approx 0.477$ which compares pretty well with the numerical simulations.

\begin{figure*}[t]
\hskip -0.4cm
\includegraphics[width=6.3cm]{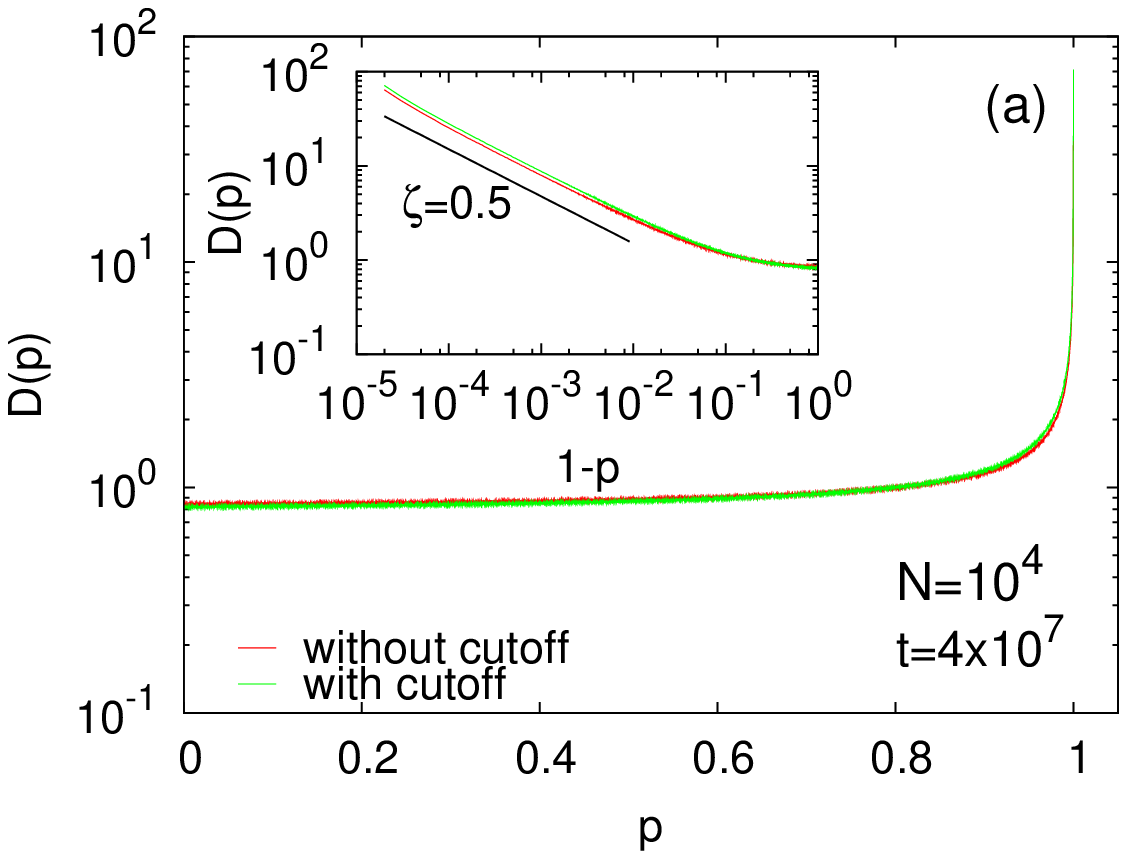} \hskip -0.4cm
\includegraphics[width=6.3cm]{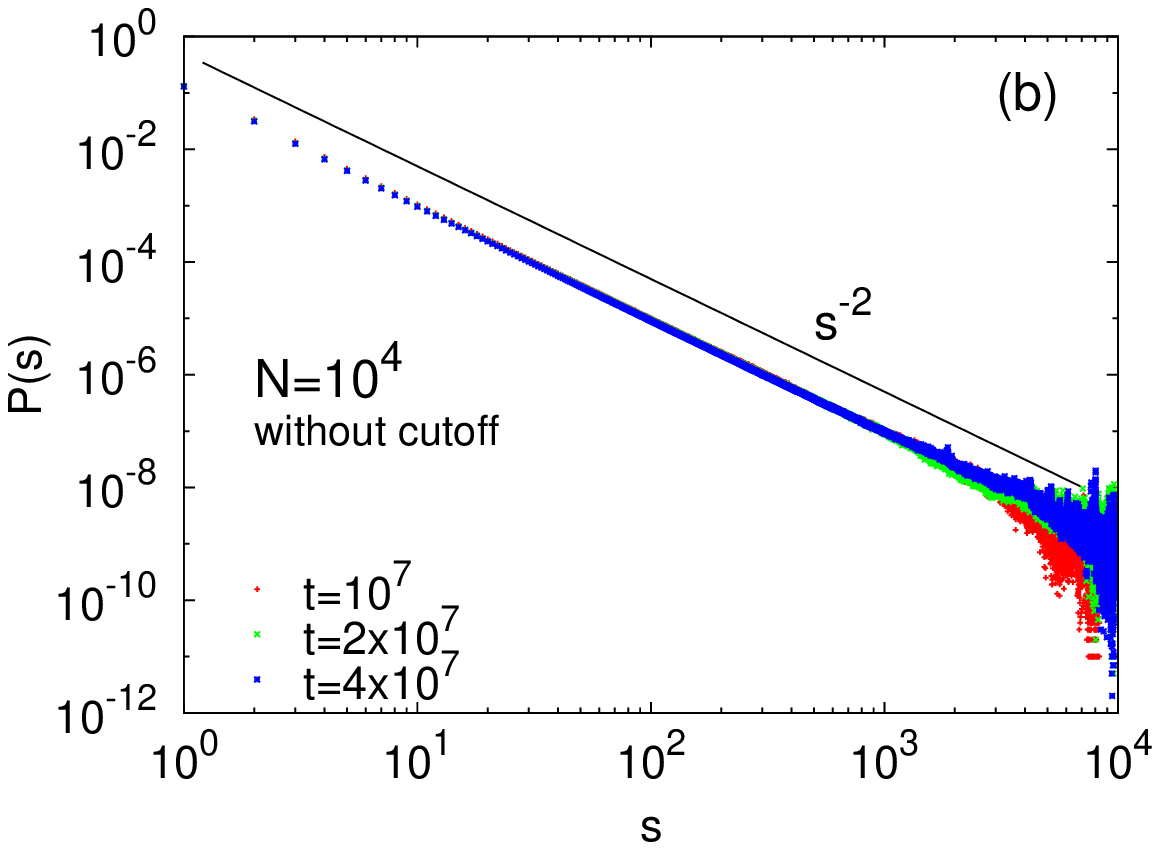} \hskip -0.4cm
\includegraphics[width=6.3cm]{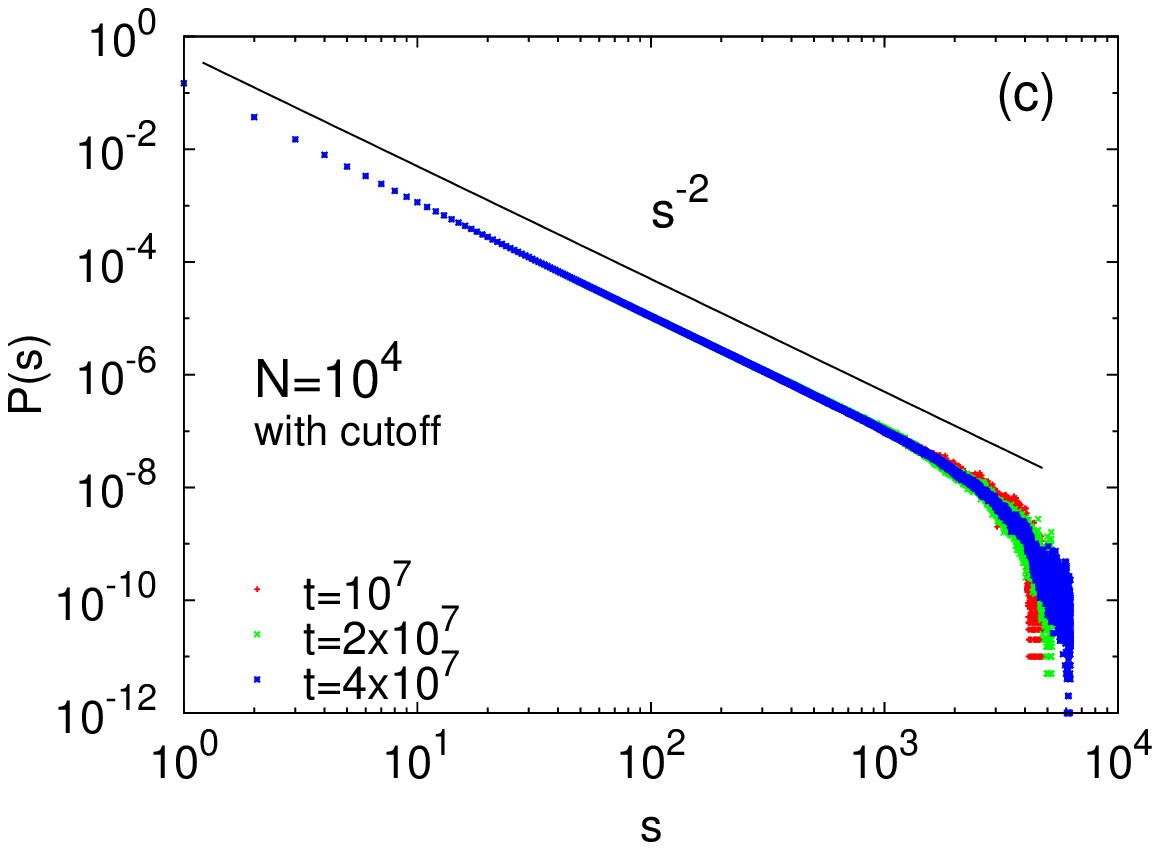}
\caption{(Color
  online) (a) The distributions $D(p)$ of $p$ for Cases I (without cutoff) and II  (with cutoff).
The inset shows $D(p)$ vs $1-p$, and the solid line is a guide to $\zeta=0.5$.
The data is shown for $N=10^4$.
(b) The distributions $P(s)$ of $s$ for  Case I (without cutoff) at different times $t=10^7,2\times10^7,4\times10^7$ for $N=10^4$.
The solid line is a guide to $\nu=2$.
(c) The distributions $P(s)$ of $s$ for  Case II  (with cutoff) at different times $t=10^7,2\times10^7,4\times10^7$ for $N=10^4$.
The solid line is a guide to $\nu=2$.
}
\label{fig:fitness}
\end{figure*}
However, when all restaurants have the same fitness ($=p$), the fraction of agents choosing restaurants
randomly is $l=1-p$, who are mobile agents, while 
 $1-l$ fraction of agents are immobile.
Then, for this mobile fraction $l$, the probability of finding a restaurant with exactly $m$ mobile agents will be
a Poissonian
\begin{equation}
 \pi(m) = \frac{l^m}{m!} \exp(-l).
\label{pimfinite}
\end{equation}
Then, we will basically have Eq.~\ref{eq:series} with $\pi(m)$ given by Eq.~\ref{pimfinite}. Thus,
\begin{eqnarray}
 1-F(l) = e^{-l} &+& (l^2 e^{-l} -l e^{-l} + l e^{-2l}) \nonumber \\
      &+& \left(\frac{l^3e^{-l}}{2} - l^2 e^{-l} + \frac{l^3 e^{-2l}}{2} + l^2e^{-2l} \right) \nonumber \\
      &+& \ldots,
\label{eq:seriesfin}
\end{eqnarray}
where $F(.)$ is the contribution to utilization from the mobile agents.
Now, the total utilization fraction will constitute of the contributions of the 
mobile and immobile agents:
\begin{equation}
 f(p)= l F(l)+(1-l)=(1-p)F(1-p)+p.
 \label{fprimep}
\end{equation}
We compute Eq.~\ref{fprimep} upto 3 terms in the series, 
and plot in Fig.~\ref{fig:util}a, and compare with numerical simulations.
In fact,  $f(p) \to 1$ as $p \to 1$,
which can easily be explained from the fact that at the limit of $p \to 1$,
there is hardly any fluctuation and $a_i(t+1)=a_i(t)$ identically.

For the case when $\Pi(p)=1$, we observe that $f$ grows with system size $N$,
roughly as $1 -b/\ln N$, which tells us that $f \to 1$ as $N \to \infty$ (Fig.~\ref{fig:util}b).
Thus, for large systems, it is possible to attain full utilization.

\subsection{Evolution with fitness}
\label{subsec:evo}
Here we apply a new strategy for the model, as follows: 
initially all the restaurants are given the same values of  $p$ and one agent per restaurant. 
Each day  agents  go to the   restaurants  obeying the rule as described in previous section 
i.e., each agent will return to the same restaurant with probability $p$ or choose any other
restaurant uniformly. By this strategy, some of the restaurants will lose agents 
and correspondingly some will gain agents compared to previous day's attendance.
Fitness plays an important role in the evolutionary models of species (see e.g., Ref.~\cite{bak1993punctuated}).
Let only the restaurants which lose agents refresh their fitness $p$ by a new  value randomly 
drawn from $\Pi(p) = (1+\delta)(1-p)^{\delta}$ in $[0,1)$ for next day.
This process may actually mean that a restaurant performing badly goes out of business
and is replaced by a new one. In the context of cities, this might mimic a process
of city decline/death and a subsequent emergence of a new city.

We study the problem for two cases: where we do not use any cutoff for $p$ (Case I) and 
where a natural cutoff in $p$ is used (Case II).

Case I: $N$ restaurants are initially assigned the same value of $p$ and  one agent in each restaurant,
and the dynamics is as described above, but the new values of $p$  are  drawn from a uniform random distribution in $[0,1)$ (i.e., $\delta=0$).
The agent distribution  $P(s)$ in the steady state follows a power law with exponent $\nu=2$. 
Also the steady state distribution $D(p)$ of $p$ in higher value of $p$ behaves as 
\begin{equation}
D(p) = \frac{A}{(1-p)^{\zeta}} + B, 
\label{eq:dp}
\end{equation}
where $A,B$ are constants and $\zeta \simeq 0.5$, as shown in Fig.~\ref{fig:fitness}a.
We checked numerically for several values of $\delta$ and find that the relation 
\begin{equation}
\zeta = \delta - \frac{1}{2}
\label{eq:zeta}
\end{equation}
holds.
Here we use $D(p)$ to distinguish from $\Pi(p)$, the former being generated out of the dynamics, while the latter
is a pre-determined distribution.

Case II: To avoid the condensation,  we use a cutoff for $p$.
For $\delta=0$ we allowed the highest value for $p$ to be $1-1/N^2$. 
We choose this cutoff since $\zeta=-1/2$ near $p=1$, which gives the cutoff to be $\epsilon=1/N^{\frac{1}{1+\zeta}}=1/N^2$.
We find that same power law  behavior with an exponential cutoff.  
Additionally, the system is ergodic; we observe that agent distribution $P_i(s)$ at any randomly selected restaurant $i$
is the same as the agents distribution computed from all restaurants (see Fig.~\ref{fig:fitness}c). 
Eq.~\ref{eq:dp} and Eq.~\ref{eq:zeta} still hold true.

\section{Empirical evidences}
\label{sec:4}
\begin{figure}[t]
\includegraphics[width=8.7cm]{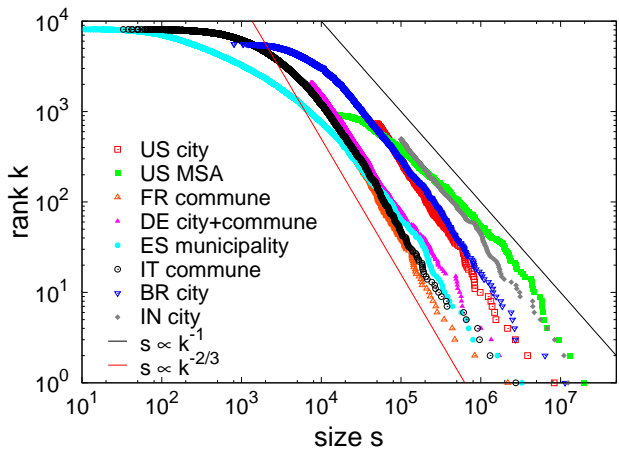}
\caption{ (Color
  online) 
Plot of size $s$ with rank $k$ of a city for different countries.
The two straight lines are respectively $s \propto k^{-1}$ and $s \propto k^{-2/3}$,
guides to the approximate extremes of the power law behavior of the data sets.
The actual exponents of the power law fits  are
given in Table.~\ref{tab:fits}.
 }
 \label{fig:dataset}
\end{figure}

In Fig.~\ref{fig:dataset}, we plot the size $s$ of cities, communes, municipalities
and their rank $k$ according to size for  several countries across the world,
and one typically observes variations in the exponents.
The slopes of the curves basically give the power law exponent $1/\gamma = \nu -1$ corresponding to the Zipf law.
We computed these exponents at the tail of the distributions using maximum likelihood estimates (MLE)~\cite{Clauset2009}
and subsequently calculated $\gamma$,  as shown in Table.~\ref{tab:fits}.

%
\begin{table}[t]
\caption{Zipf exponents for different countries, computed using MLE.
For USA, we used two data sets: cities with population above $50,000$,
and for Metropolitan Statistical Area (MSA). 
For India, data for cities with population above $100,000$ are used.}
\begin{tabular}{|l|l|l|c|c|}
\hline
Country & Year & demarcation  & $\gamma$ \\
\hline
USA~\cite{US1} & 2012 & city population $> 50,000$   &  0.74(2)   \\
USA~\cite{US1} & 2012 & MSA & 0.91(2)   \\
France~\cite{France} & 2006 & commune   &   0.67(1)  \\
Germany~\cite{Germany} & 2011 & city \& commune   & 0.85(2)    \\
Spain~\cite{Spain} & 2011 & municipality  &  0.77(1)   \\
Italy~\cite{Italy} & 2010 & commune   &  0.77(1)  \\
Brasil~\cite{Brasil} & 2012 & city   &  0.88(1)  \\
India~\cite{India} & 2011 & city population $> 100,000$  & 0.63(1)   \\
\hline
\end{tabular}
\label{tab:fits}
\end{table}

\section{Summary and discussions}
\label{sec:5}
The social and economic reasons for the development  of an urban agglomeration 
or a city~\cite{batty2008size} involve growth over time as well as migration, 
decay, as well as death, due to natural or  economic (industrial) reasons.
In this article we model city growth as a resource utilization problem,
specifically in the context of city size distributions.
Zipf law for city size distribution can be thought to be a consequence of the variation
in the quality of available services, which can be measured in terms of various amenities.
We argue that this measure can be characterized by an intrinsic fitness. 
We make a correspondence from the population in cities to the number of customers in 
restaurants in the framework of the Kolkata Paise Restaurant
problem, where each restaurant is characterized by an intrinsic fitness $p$
similar to the difference in the quality of services in different cities.
The basic model is introduced in Sec.~\ref{sec:2}.
In Sec.~\ref{subsec:size}, we calculate the size distributions,
and in Sec.~\ref{subsec:util}, the
exact value of the utilization fraction for  
the case when choices are made independent of fitness.
Results for the case with uniform fitness are also reported there.
When fitness is uniformly distributed, it can give rise to a power (Zipf) law
for the number of customers in each restaurant.
We investigate a variant of the model (Sec.~\ref{subsec:evo}) where the fitness can be seen
as the ability to stay in the business. When a restaurant loses customers, 
its fitness is refreshed with another random value. 
In the steady state, the power-law distribution of the number of customers still holds, 
implying the robustness of the model (with fitness distribution characterized by power laws).
Using a simple mechanism in which agents compete for available resources,
and find the best solution using iterative learning, we show that the emergent 
size distribution given by the number of customers in restaurants is a power law.
It may be noted that even though we consider here the particular case of $N=R$, the
possibility that the restaurants (cities) adjust (learn) their fitness according to the
past experience, induce the power law distribution of the customers (Sec.~\ref{subsec:evo}),
leaving many restaurants (cities) vacant or dead.

Although our model, using a very simple mechanism of migration of agents
in a system of cities (restaurants) with a random fitness distribution
reproduces the Zipf law, we have not taken into consideration 
any spatial structure, the costs
incurred in migration/transport of agents (cf. Ref.~\cite{gastner2006spatial}),
and the spatial organization of facilities~\cite{um2009scaling}
which may emerge as important factors that drive the flow of population from
one city to another. We did not incorporate several details 
of social organization but kept the  bare essential ingredients
that can give rise to Zipf law.
Although our study limits to a mean field scenario, being defined on 
a regular, fully connected network, one can as well study 
the problem on directed networks~\cite{chatterjee2009kinetic}
which takes into account the asymmetry in the flows between different nodes (cities).

\begin{acknowledgments}
B.K.C. and A.C. acknowledges support from B.K.C.'s J.~C.~Bose Fellowship and  Research Grant.
We also thank the anonymous referee for independent numerical checking and confirming some of our
crucial results and for suggesting inclusion of some detailed calculations in 
Sections \ref{subsec:size}, \ref{subsec:util} and \ref{subsec:evo} for the benefit of the readers.

\end{acknowledgments}

\bibliographystyle{h-physrev3}
\bibliography{kpr_zipf.bib}

\end{document}